\documentclass[aps,twocolumn,showpacs,showkeys]{revtex4}
\usepackage{graphicx}
\usepackage{amssymb}
\usepackage{bm}
\begin{document}
\setcounter{page}{1}
\title{Interacting Particle Systems in Complex Networks}
\author{Jae Dong \surname{Noh}}
\email{jdnoh@uos.ac.kr}
\affiliation{Department of Physics, University of Seoul, Seoul
130-743, Korea}
\date{}

\begin{abstract}
We present our recent work on stochastic particle systems on complex networks. 
As a noninteracting system we first consider 
the diffusive motion of a random walker on heterogeneous complex networks.
We find that the random walker is attracted toward nodes with 
larger degree and that the random walk motion is asymmetric.
The asymmetry can be quantified with the random walk centrality, the
matrix formulation of which is presented. As an interacting system, we 
consider the zero-range process on complex networks. We find that a
structural heterogeneity can lead to a condensation phenomenon.
These studies show that structural heterogeneity plays an important
role in understanding the properties of dynamical systems.
\end{abstract}



\maketitle

\section{INTRODUCTION}
During the last several years, complex networks have been attracting a lot of
interest in the statistical physics
community~\cite{Watts98,Albert02,Dorogovtsev02,Newman03}. Complex networks
are characterized by a heterogeneous structure, which is
accounted for by a broad degree distribution. The degree distribution
$P_{deg}(k)$
denotes the fraction of nodes with degree $k$, where the degree of a node
means the number of links attached to it.
In contrast to periodic regular lattices, many real-world complex networks 
exhibit fat-tailed degree
distributions. Particularly, some networks displays
the power-law degree distribution
$P_{deg}(k) \sim k^{-\gamma}$
with the degree exponent $\gamma$. Such networks are called 
scale-free~(SF) network. Examples include the 
Internet, the World Wide Web, scientific collaboration networks, and so on.

Recent studies have revealed that the heterogeneous structure of complex 
networks has a nontrivial effect on physical problems defined upon them. 
In the percolation problem, for instance, the percolation threshold vanishes
as the second moment of the degree $\langle k^2\rangle$  
diverges~\cite{Cohen00}. Hence, 
a SF network with $\gamma\leq 3$ has a percolating cluster at any
finite node/edge density. Similarly, the onsets of phase transitions in
(non-)equilibrium systems strongly depend on the heterogeneity through
$\langle k^2\rangle$~\cite{Pastor-Satorras01,Goltsev03}. 
The scaling behavior of dynamical systems is also influenced by the
structures 
of underlying complex networks~\cite{Gallos04,Catanzaro05,Noh06}.

In this paper we present a brief review of recent works on 
stochastic particle systems on complex networks~\cite{Noh04,Noh05,Noh05b}.
This is to understand how the structural heterogeneity 
influences the dynamical and the stationary-state properties of particle 
systems~\cite{citea,citeb,citec,cited}.
We address this issue in the context of the random walk process and 
the zero-range process, which will be presented in Sec.~\ref{Sec2} and
Sec.~\ref{Sec3}, respectively. 

This paper is organized as follows.
In Sec.~\ref{Sec2}, we consider a random walk on general networks.
We are particularly interested in the stationary-state probability 
distribution for the random walker position and the mean first passage 
time~(MFPT). The random walk study shows that a diffusing particle tends to be
attracted toward nodes with large degree. It also shows that the diffusive
motion on complex networks is asymmetric in that the MFPT from one node to
another is different from that in the reverse direction. The asymmetry
can be quantified by the so-called random-walk centrality. We present a
matrix formulation for the quantity.
The random-walk study hints that condensation will occur in interacting 
many-particle systems on complex networks. 
In Sec.~\ref{Sec3}, we consider the zero-range process on complex networks.
We will show 
that the structural heterogeneity of underlying networks leads to
condensation in which a macroscopic finite fraction of total particles 
is concentrated on larger degree nodes. 
We also present a criterion for the condensation.
We summarize the paper in Sec.~\ref{Sec4}

\section{Random walk}\label{Sec2}
We introduce our notations. The total numbers of nodes and
edges in a network are denoted by $N$ and $L$, respectively. 
The connectivity of a
network is represented with the adjacency matrix $\mathbf{A}$ 
whose matrix elements $a_{ij}~(i,j=1,\ldots,N)$ take the values of $1~(0)$ 
if there is an~(no) edge between two nodes $i$ and $j$. 
The adjacency matrix is assumed
to be symmetric~($a_{ij}=a_{ji}$); that is, we only consider undirected
networks.  The degree of a node $i$ is denoted by $k_i$, which is given by 
$k_i = \sum_j a_{ij}$.

At each time step, a random walker performs a jump from a node to one 
of its neighboring nodes selected randomly.
Let us denote the probability to find the walker 
at each node $i$ at the $n$th time step by $p_i(n)$. The time evolution 
for the probability distribution is given by
\begin{equation}
\mathbf{P}(n+1) = \mathbf{W} \mathbf{P}(n) ,
\end{equation}
where $\mathbf{P}(n) \equiv (p_1(n),\ldots,p_N(n))^t$ is a column vector 
and $\mathbf{W}$ is the transition matrix with matrix element $W_{ij}
\equiv a_{ij}/k_j$.~(The superscript $^t$ denote the transpose.) 
With an initial condition $\mathbf{P}(0)$, the 
probability distribution at arbitrary time step is given by 
$\mathbf{P}(n) = \mathbf{W}^n \mathbf{P}(0)$. In particular, the transition
probability of the walker from a node $j$ to $i$ in $n$ steps is given by
$p_{ij}(n) = (\mathbf{W}^n)_{ij}$~\cite{comment0}.

In Ref.~\cite{Noh04}, we showed that the stationary-state probability 
is given by $\mathbf{P}(\infty) = (p_1(\infty),\ldots,p_N(\infty))^t$
with 
\begin{equation}
p_i(\infty) = k_i / (2 L) . 
\end{equation}
Actually, one can check easily that 
$\mathbf{P}(\infty)$ is the right eigenvector of $\mathbf{W}$ with the 
eigenvalue 1. The corresponding left eigenvector is the row vector 
$\mathbf{1} = (1,\ldots,1)$. The result states that the frequency with which 
the walker visits a node is directly proportional to its degree.
Therefore, in a heterogeneous network, the walker is attracted toward larger
degree nodes, which suggests that condensation or jamming can occur in 
many-particle systems, which will be discussed in next section.

The dynamical aspect of the random walk is studied with the MFPT.
We showed in Ref.~\cite{Noh04} that the MFPT $T_{ij}$ from a node 
$j$ to $i$ is given by
\begin{equation}\label{MFPT}
T_{ij} = \frac{2L}{k_i} \left[ R^{(0)}_{ii} - R^{(0)}_{ij}\right]
(1-\delta_{ij}) + \frac{2L}{k_i} \delta_{ij} \ ,
\end{equation}
where $\delta_{ij}$ is the Kronecker delta symbol and 
\begin{equation}\label{R0}
R^{(0)}_{ij} \equiv \sum_{n=0}^\infty \left\{ p_{ij}(n) -
p_i(\infty)\right\} \ .
\end{equation}
Moreover, we found that the MFPT satisfies the relation
\begin{equation}\label{TijTji}
T_{ij} - T_{ji} = 1/C_i - 1/C_j \ ,
\end{equation}
where $C_i$ is the random walk centrality~(RWC) defined as
\begin{equation}\label{RWC}
C_i \equiv \frac{k_i}{2L R^{(0)}_{ii}} \ .
\end{equation}
The result shows that the random walk speed to a node is determined by the
RWC: The larger the RWC a node has, the faster the random walk motion 
to it is. 

For the MFPT and the RWC, one needs to evaluate the matrix $\mathbf{R}^{(0)}$
with the element  $R^{(0)}_{ij}$ defined in Eq.~(\ref{R0}). 
We present the matrix formulation for it.
First, we define a matrix $\mathbf{V} \equiv \mathbf{P}(\infty)
\mathbf{1}$ by using the product of the column vector $\mathbf{P}(\infty)$ and
the row vector $\mathbf{1}$, which are the right and the left eigenvectors
of $\mathbf{W}$, respectively, with the eigenvalue $1$. 
It is easy to check that 
the matrix element is given by $V_{ij} = p_i(\infty)$. Hence, one can write
that $R^{(0)}_{ij} = \sum_{n=0}^{\infty} \left( \mathbf{W}^n - \mathbf{V}
\right)_{ij}$ or 
\begin{equation}
\mathbf{R}^{(0)} = \sum_{n=0}^\infty ( \mathbf{W}^n - \mathbf{V})\ .
\end{equation}
Note that the matrix $\mathbf{V}$ is the projection of $\mathbf{W}$ onto the
subspace with the eigenvalue $1$, which yields the relations
$\mathbf{V}^n = \mathbf{V}$ for $n>0$ and $\mathbf{W} \mathbf {V} =
\mathbf{V} \mathbf{W} = \mathbf{V}$. Combining these, one can find easily
that $\mathbf{W}^n - \mathbf{V}$ is equal to $(\mathbf{W}-\mathbf{V})^n$ 
for $n>0$ and $(\mathbf{I} - \mathbf{V})$ for $n=0$.
Consequently, the matrix $\mathbf{R}^{(0)}$ is given by
\begin{equation}\label{R0_matrix}
\mathbf{R}^{(0)} = \left[ \mathbf{I} - \left( \mathbf{W} - \mathbf{V} 
\right) \right]^{-1} - \mathbf{V} \ .
\end{equation}
The MFPT and the RWC are then calculated from Eqs.~(\ref{MFPT}) and 
(\ref{RWC}), respectively.

\begin{figure}[t]
\includegraphics[width=0.7\columnwidth]{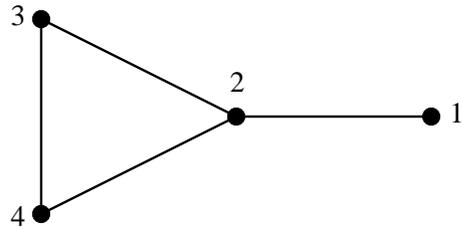}
\caption{Model network with $N=4$ nodes and $L=4$ edges.} \label{fig1}
\end{figure}

We demonstrate the way one calculates the MFPT and RWC by using the matrix
$\mathbf{R}^{(0)}$ with the simple example network shown in Fig.~\ref{fig1}.
For this network, the random walk transition matrix is given by
\begin{equation}
\mathbf{W} = \left(
\begin{array}{cccc}
0 & 1/3 & 0   & 0   \\
1 & 0   & 1/2 & 1/2 \\
0 & 1/3 & 0   & 1/2 \\
0 & 1/3 & 1/2 & 0   
\end{array} \right) \ .
\end{equation}
The matrix $\mathbf{V}$ is given by $\mathbf{P}(\infty) (1,1,1,1)$ with
the stationary-state probability $\mathbf{P}(\infty) = (1/8,3/8,1/4,1/4)^t$.
Thus, using Eq.~(\ref{R0_matrix}), one obtains the matrix 
$\mathbf{R}^{(0)}$:
\begin{equation}
\mathbf{R}^{(0)} = \left(
\begin{array}{rrrr}
57/64 & 1/64  & -15/64 & -15/64 \\
3/64  & 27/64 & -21/64 & -21/64 \\
-15/32 & -7/32 & 59/96 & -5/96  \\
-15/32 & -7/32 & -5/96  & 59/96
\end{array} \right) \ .
\end{equation}
With the matrix, one can easily obtain the MFPT by using Eq.~(\ref{MFPT}). 
The resulting MFPT's are given by
\begin{equation}
(T_{ij}) = \left(
\begin{array}{cccc}
8    &   7   &   9   &  9   \\
1    & 8/3   &   2   &  2   \\
13/3 & 10/3  &   4   &  8/3 \\
13/3 & 10/3  &  8/3  &  4   \\
\end{array} \right) \ .
\end{equation}
This shows that  $T_{ij} \neq T_{ji}$, the asymmetry of the random 
walk motion. Using Eq.~(\ref{RWC}), we find that the RWC is given by
$C_1=8/57$, $C_2 = 8/9$, and $C_3=C_4 = 24/59$.
In this example, node $2$ has the largest RWC, so one can say that 
node $2$ is the most important node in the diffusion process.

\section{Zero-range process}\label{Sec3}
The study in the previous section shows that the structural heterogeneity of
underlying networks leads to the non-uniform particle distribution, which
suggests that the interaction may play an important role in
many-particle systems in heterogeneous networks. 
We study the effect of the
heterogeneity on the property of interacting particle systems by using 
the zero-range process~(ZRP). The ZRP is a
driven diffusive particle system and has been studied extensively on regular 
lattices~\cite{Evans00,Krug00,Evans05}.  
We extend those studies to the heterogeneous network.

The ZRP is defined as follows:
Consider an undirected network of $N$ nodes and $L$ edges with the degree
distribution $P_{deg}(k)$.
There are $M$ particles on the network with the density $\rho=M/N$. 
We denote the
number of particles at each node $i$ as $m_i=1,2,\ldots$. 
Those particles move
around the network according to the dynamic rule: At each node $i$, one
particle jumps out of it at a given rate $q_i(m_i)$ and then moves to one
of the neighboring nodes selected randomly.

Interestingly, the stationary-state probability distribution is known
exactly for the ZRP~\cite{Evans05}. It is given in
factorized form as
\begin{equation}\label{Pzrp}
P(m_1,\ldots,m_N) = \frac{1}{Z}\prod_{i=1}^N f_i(m_i) 
\end{equation}
where $Z$ is a normalization factor and
\begin{equation}
f_i(m) = \prod_{m'=0}^m \left( \frac{p_i(\infty)}{q_i(m')} \right)
\end{equation}
for $m>0$ and $f_i(m=0)=1$.
Here, $p_i(\infty)$ is the 
stationary-state probability for a single-particle random-walk problem 
on the network. According to the result in the previous
section, we have $p_i(\infty)=k_i/(2L)$. 
Using the probability distribution in Eq.~(\ref{Pzrp}), 
one can calculate the mean value of the
particle number at each node. It is given by
\begin{equation}
\langle m_i \rangle = z \frac{\partial}{\partial z} \ln F_i (z) \ ,
\end{equation}
where $F(z) \equiv \sum_{m=0}^\infty z^m f_i(m)$. The fugacity variable $z$
is limited to the interval $z<z_c$, where $z_c$ is given by 
the radius of convergence of the series and
should be determined from the self-consistency equation $M = \sum_{i}\langle
m_i \rangle$.

The particle interaction can be incorporated into the system through 
the jumping rate function $q_i(m)$. In order to stress the effect of the
structural heterogeneity, we concentrate on the simplest case with a
constant jumping rate function $q_i(m)=1$. 
This is a special case of the model studied in Refs.~\cite{Noh05} and 
\cite{Noh05b}.
For this case, the function $F_i(z)$ is given by 
$F_i(z) = \sum_{m=0}^\infty (z k_i)^m = 1 / (1 - zk_i)$~\cite{comment}, and 
the mean occupation number is given by 
\begin{equation}
\langle m_i \rangle = \frac{ z  k_i } { 1 - z k_i } \ .
\end{equation}
The fugacity is limited within the interval $z<z_c = 1/k_M$, where $k_M$ is
the maximum degree, and its value should be determined from the
self-consistency equation
\begin{equation}
\rho = \frac{1}{N} \sum_i \frac{zk_i}{1-zk_i} = 
\sum_k \frac{zk P_{deg}(k)}{1-zk} \ .
\end{equation}

Note that the occupation number is a monotonically increasing function of
the degree. Hence, nodes with the maximum degree $k_M$ have the largest
occupation number, which could be divergent~(macroscopically large) 
in the $z\to z_c=1/k_M$ limit. This suggests a possibility that 
heterogeneity-induced condensation may take place. Condensation refers
to a phenomenon in which some nodes are occupied by macroscopically many
numbers of particles.

The occupation number distribution depends strongly on 
the degree distribution $P_{deg}(k)$. 
We demonstrate the condition for heterogeneity-induced condensation
in the following:
(i) Suppose that the degree distribution $P_{deg}(k)$ is supported on a finite
interval $k\leq k_M$, where the maximum degree $k_M$ is finite. 
The general feature of this case 
can be understood easily with an explicit example described by 
\begin{equation}\label{Pk1_k2}
P_{deg}(k) = a \delta_{k,k_1} + b \delta_{k,k_2} \ , 
\end{equation}
where $a+b=1$ and $k_1 < k_2 = k_M<\infty$. Periodic lattices have a
degree distribution of this form. For finite $a$ and $b$, it is easy to
verify that the self-consistency equation $\rho = azk_1/(1-zk_1) + b
zk_2/(1-zk_2)$ has the solution $z$ with finite $\epsilon=1/k_2 - z$ for all
values of $\rho$. Since $\epsilon$ is finite, the mean occupation number at 
each node is either $zk_1/(1-zk_1)$ or $zk_2/(1-zk_2)$ and finite, so the
system does not display condensation.

The situation changes drastically when a heterogeneity is introduced 
in the degree distribution.
Let us assume that there is an impurity node having a higher degree than the
other nodes. Such networks can be described by a degree distribution 
of the type as in Eq.~(\ref{Pk1_k2}) with $b=1/N$. The nodes with degree
$k_1$ and $k_2$ will be called the normal and the impurity nodes,
respectively.
The self-consistency equation becomes 
\begin{equation}
\rho = \left(1-\frac{1}{N}\right) \frac{z k_1}{ 1-z k_1} +  \frac{1}{N}
\frac{z k_2}{1-z k_2} \ .
\end{equation}
Since $z<z_c=1/k_2$, the mean occupation number at the normal nodes 
is bounded above by $\rho_c \equiv \left. zk_1/(1-zk_1) 
\right|_{z=z_c} = k_1/(k_2-k_1)$, 
which implies that the whole fraction of particles cannot be 
accommodated into the normal nodes when $\rho>\rho_c$. 
Consequently, in the large-$N$ limit, one finds that the 
occupation number is given by $m_{nor.}=\rho_c$ for normal nodes 
and by $m_{imp.}=N(\rho-\rho_c)$ for the impurity node. That is to say, the
system displays condensation when the particle density exceeds the
threshold value. 

These considerations using the model degree distribution given in
Eq.~(\ref{Pk1_k2}) can be easily extended to the general case: 
If $P_{deg}(k)$ is supported on a finite interval $k\leq k_M$ and
$P_{deg}(k=k_M)$ is finite, condensation 
does not occur at any value of $\rho$. On the other hand, 
if there are a few nodes with a higher degree than the other 
nodes~($P_{deg}(k=k_M) = \mathcal{O}(1/N)$),
heterogeneity-induced condensation can take place. It occurs 
when the particle density
exceeds a threshold value $\rho_c = \sum_{k<k_M} k P_{deg}(k) / (k_M - k)$.

(ii) We now consider a degree distribution, which is supported on the
unbounded interval of $k$ in the limit of $N\to\infty$. 
The SF networks with
the power-law degree distribution $P_{deg}(k)\sim k^{-\gamma}$ and the
Erd\H{o}s-R\'enyi random networks with the
Poisson distribution $P_{deg}(k) = e^{-\langle k\rangle} \frac{\langle
k\rangle^k}{k!}$ belong to this class. The ZRP's in the SF
networks and the random networks are studied in Refs.~\cite{Noh05} and
\cite{Kim05}. 
For finite $N$, those networks still have a finite maximum degree
$k_M=k_M(N)$, which will diverge as $N\to\infty$.
For example, $k_M \sim N^{1/(\gamma-1)}$ and $k_M \sim \ln N$ for the SF
networks and the random networks, respectively.

One can write the self-consistency equation in the form of $\rho = \rho_n
+\rho_s$ where $\rho_n \equiv \sum_{k<k_M}zkP_{deg}(k)/(1-zk)$ is the 
fraction of particles in nodes with $k<k_M$ and $\rho_s \equiv 
zk_M P_{deg}(k_M) /(1-zk_M)$ is the fraction of particles at nodes 
with $k=k_M$.
In the large-$N$ limit, one can make an approximation $\rho_n \simeq
\int^{k_M}dk~ zkP_{deg}(k)/(1-zk)$. The integral interval can be divided
into two parts,  $k\ll k_M$ and $k\sim k_M$. The first term makes an
$\mathcal{O}(z)$ contribution while the second term can make a singular
contribution as $z\to z_c=1/k_M$.
Thus, for $z=z_c-\epsilon$ with small $\epsilon$, 
the quantity $\rho_n$ can be approximated as
\begin{eqnarray*}
\rho_n &\simeq& k_M P_{deg}(k_M) \int^{k_M} dk z / (1-zk) +
\mathcal{O}(1/k_M)\\
       &\simeq& k_M P_{deg}(k_M) \left| \ln (1-z k_M)\right|
+\mathcal{O}(1/k_M) \\
       &\simeq& k_M P_{deg}(k_M) \left| \ln \epsilon k_M\right|
+\mathcal{O}(1/k_M) .
\end{eqnarray*}
Since $k_M\to \infty$ in the $N\to\infty$ limit, we can neglect the second
term coming from the integral in the interval $k\ll k_M$.
For networks with finite mean degree, the degree distribution should decay
faster than $P_{deg}(k) \sim k^{-2}$. This insures that the first term also
vanishes in the $k_M\to\infty$ limit for any finite value of $\epsilon$. 
This analysis shows that the whole fraction of particles in the nodes with
$k<k_M$ vanishes in the $N\to\infty$ limit, 
which implies that the whole fraction
of particles should be condensed into the node with the maximum degree
$k=k_M$. 
Therefore, we conclude that 
heterogeneity-induced condensation always occur at any value of $\rho$ on
networks with a degree distribution supported on the infinite interval.

\section{Summary}\label{Sec4}
We have considered stochastic particle systems on complex networks and
studied the effect of the structural heterogeneity in networks on
the properties of the particle systems. In the random-walk model, we 
showed that the heterogeneity makes the random walk motion biased and 
asymmetric:
The random walker visits higher degree nodes more frequently, and the
diffusion to higher random walk centrality nodes is faster than it is to lower 
random walk centrality nodes. We have presented the matrix formalism for the
mean first passage time and the random walk centrality. In the ZRP
with a constant jumping rate, we showed that the structural 
heterogeneity can give rise to condensation. For networks with a
finite maximum degree in the $N\to\infty$ limit, the condensation occurs
only when there are a few impurity nodes with higher degree and the particle
density is greater than a threshold value. In that case, the impurity nodes
are occupied by a macroscopically large number of particles, and the other
nodes by a finite number of particles. For networks with a 
diverging maximum degree in the $N\to\infty$ limit, 
condensation always occurs at any finite value of the
particle density. In this case, the whole fraction of particles is
condensed onto the node with maximum degree.

\begin{acknowledgments}
This work was supported by the Korea Research Foundation Grant
(KRF-2004-041-C00139). 
\end{acknowledgments}


\begin{references}
\bibitem{Watts98} D. J. Watts and S. H. Strogatz, 
        Nature (London) {\bf 393}, 440 (1998).
\bibitem{Albert02} R. Albert and A.-L. Barab\'asi,
        Rev. Mod. Phys. {\bf 74}, 47 (2002).
\bibitem{Dorogovtsev02} S. N. Dorogovtsev and J. F. F. Mendes, 
        Adv. Phys. {\bf 51}, 1079 (2002).
\bibitem{Newman03} M. E. J. Newman, SIAM Rev. {\bf 45}, 167 (2003).
\bibitem{Cohen00} R. Cohen, K. Erez, D. ben-Avraham, and S. Havlin,
        Phys. Rev. Lett. {\bf 85}, 4626 (2000).
\bibitem{Pastor-Satorras01} R. Pastor-Satorras and A. Vespignani,
        Phys. Rev. Lett. {\bf 86}, 3200 (2001).
\bibitem{Goltsev03}  A. V. Goltsev, S. N. Dorogovtsev, and J. F. F. Mendes,
        Phys. Rev. E {\bf 67}, 026123 (2003).
\bibitem{Gallos04} L. K. Gallos and P. Argyrakis,
        Phys. Rev. Lett. {\bf 92}, 138301 (2004).
\bibitem{Catanzaro05} M. Catanzaro, M. Bogu\~na, and R. Pastor-Satorras,
        Phys. Rev. E {\bf 71}, 056104 (2005).
\bibitem{Noh06} J. D. Noh and S.-W. Kim,
        J. Korean Phys. Soc. {\bf 48}, S202 (2006).
\bibitem{Noh04} J. D. Noh and H. Rieger,
        Phys. Rev. Lett. {\bf 92}, 118701 (2004).
\bibitem{Noh05} J. D. Noh, G. M. Shim, and H. Lee,
        Phys. Rev. Lett. {\bf 94}, 198701 (2005).
\bibitem{Noh05b} J. D. Noh, Phys. Rev. E {\bf 72}, 056123 (2005). 
\bibitem{citea} K.-I. Goh, B. Kahng, and D. Kim, J. Korean Phys. Soc. {\bf
46}, 551 (2005).
\bibitem{citeb} B. J. Kim, J. Korean Phys. Soc. {\bf 46}, 722 (2005).
\bibitem{citec} K. Rho, B. Kahng, and D. Kim, J. Korean Phys. Soc. {\bf 47},
558 (2005).
\bibitem{cited} S. H. Lee and H. Jeong, J. Korean Phys. Soc. {\bf 48}, 186
(2006).
\bibitem{comment0} Notice that the subscript notation in this paper is
different from that in Ref.~\cite{Noh04}.
\bibitem{Evans00} M. R. Evans, Braz. J. Phys. {\bf 30}, 42 (2000).
\bibitem{Krug00} J. Krug, Braz. J. Phys. {\bf 30}, 97 (2000).
\bibitem{Evans05} M. R. Evans and T. Hanney,
         J. Phys. A {\bf 38}, R195 (2005).
\bibitem{comment} The global $2L$ factor in $p_i(\infty)=k_i/(2L)$ can be
        neglected because it amounts to a redefining of the fugacity $z$.
\bibitem{Kim05} S. Kwon, S. Lee, and Y. Kim, Phys. Rev. E {\bf 73}, 056102 (2006).
\end{references}
\end{document}